\documentclass[aps, prl, floatfix,  twocolumn, amsmath, superscriptaddress, showpacs]{revtex4-1} 
\usepackage{graphicx}  
\usepackage{amsfonts}
\usepackage{mathtools}
\usepackage{color}
\usepackage{xr}
\DeclarePairedDelimiter{\ceil}{\lceil}{\rceil}
\DeclarePairedDelimiter{\floor}{\lfloor}{\rfloor}

\newcommand{\veff}{v_{\textrm{eff}}}  

\newcommand{\PRLsec}[1]{\noindent\textit{#1}---}
\newcommand{\rd}{r_{\Delta}} 
\newcommand{\dt}{\Delta t} 
\newcommand{\dx}{\Delta x} 

\begin{document}

\title{Symmetry breaking in optimal timing of traffic signals on an idealized two-way street} 

\author{Mark J. Panaggio}
\email[email: ]{markpanaggio2014@u.northwestern.edu}
\affiliation{Department of Engineering Sciences and Applied Mathematics, Northwestern University, Evanston, IL 60208, USA}

\author{Bertand J. Ottino-L\"offler}
\affiliation{Department of Mathematics, California Institute of Technology, Pasadena, CA 91126, USA}

\author{Peiguang Hu}
\affiliation{Engineering Science Program, National University of Singapore, Singapore}

\author{Daniel M. Abrams}
\affiliation{Department of Engineering Sciences and Applied Mathematics, Northwestern University, Evanston, IL 60208, USA}
\affiliation{Northwestern Institute on Complex Systems, Northwestern University, Evanston, IL 60208, USA}
\date{ \today}

\begin{abstract}
Simple physical models based on fluid mechanics have long been used to understand the flow of vehicular traffic on freeways; analytically tractable models of flow on an urban grid, however, have not been as extensively explored. In an ideal world, traffic signals would be timed such that consecutive lights turned green just as vehicles arrived, eliminating the need to stop at each block. Unfortunately, this ``green wave'' scenario is generally unworkable due to frustration imposed by competing demands of traffic moving in different directions.  Until now this has typically been resolved by numerical simulation and optimization.  Here, we develop a theory for the flow in an idealized system consisting of a long two-way road with periodic intersections.  We show that optimal signal timing can be understood analytically and that there are counter-intuitive asymmetric solutions to this signal coordination problem.  We further explore how these theoretical solutions degrade as traffic conditions vary and automotive density increases.
\end{abstract}
\pacs{89.40.Bb,89.75.-k,64.60.Cn}
\maketitle

\PRLsec{Introduction}The physics of traffic flow has been studied for more than half a century \cite{Lighthill1955,Richards1956,Webster1958,Morgan1964,Newell1967,Chowdhury2000,Kerner2004}. On freeways, traffic has been successfully modeled as a nonlinear fluid making analytical solution possible \cite{Richards1956}.  On urban grids, however, the nonlinear effects of timed traffic signals make most models analytically intractable \cite{Bavarez1967,Toledo2004,Gershenson2005,Lammer2008,Kerner2013}.  

The inefficient timing of traffic signals is responsible for up to 10\% of traffic delays \cite{NTSRC2012}. These delays cause commuters to waste dozens of hours in traffic each year, leading to billions of dollars in wasted fuel and a large environmental cost \cite{Schrank2012}. Coordination between traffic signals has proven to be a cost effective way to reduce these delays dramatically \cite{NTSRC2012}.

Signal timing schemes fall into two categories: real-time and pre-timed \cite{Webster1958,Koonce2008}.  Real-time schemes make adaptive use of information about traffic density and localized conditions to trigger light cycle changes \cite{Robertson1991}.  Unfortunately, this information is not readily available at most intersections, and installing the necessary detectors can be prohibitively expensive \cite{Yin2008}.

Pre-timed schemes employ detailed computer simulation and heuristic optimization tools such as genetic algorithms \cite{Abdelfatah1998,AbuLebdeh2000,Ceylan2004} to search for optimal timings \cite{Robertson1969,Yin2008}. Once a scheme is generated, it can be relatively inexpensive to implement, but generating such a scheme requires computational resources that are often beyond the capacity of local government. Where traffic demands fluctuate, these schemes can quickly become outdated, so there is no guarantee that they will be optimal by the time they are implemented \cite{Robertson1991}.
 
\PRLsec{Motivation}Many theoretical questions about optimal traffic signal timing remain unanswered.  For a single one-way street, the best solution is a so-called \textit{green wave} \cite{Morgan1964,Brockfeld2001}, in which vehicles leaving a light at the instant it turns green arrive at all subsequent lights at the instant they turn green \cite{Bavarez1967}.  In theory, this means that vehicles traveling at the speed limit will never stop at a red light, though in practice this fails when traffic density exceeds a ``jamming threshold'' \cite{Kerner2013}.  

It is impossible to achieve a bidirectional green wave on an arbitrary two-way street due to the inherent frustration of competing demands in each direction \cite{Toledo2007}.  Past theoretical work has focused on maximizing the ``bandwidth'' \cite{Morgan1964,Newell1967,Robertson1969,Robertson1991,Gartner1991}---the interval of time in which vehicles can progress through all traffic signals without stopping---of a finite segment of road. There are several drawbacks to this approach.  First of all, bandwidth is not a direct measure of efficiency, so the solution that maximizes bandwidth may not minimize total trip time, stops or delay \cite{Bavarez1967}.  Secondly, for long roads, non-zero bandwidth is often unachievable in one direction, making this approach incapable of improving upon one-way schemes without arbitrarily dividing the road into subsections.  

\PRLsec{Our model}We consider a highly simplified model of an infinite two-way street with traffic lights along the entire length \cite{Nagatani2006}. We fix the spacing between consecutive lights $\dx$ and set $x_n=n\dx$, where $x_n$ denotes the position of light $n$.

We let $\phi_n(t)$ denote the phase of light $n$ at time $t$, taking light $n$ to be green when 
\[ 
  2N\pi \leq \phi_n(t) < (2N+1)\pi
\] 
and red when 
\[
  (2N+1)\pi \leq \phi_n(t) < (2N+2)\pi~,
\] 
where N is any integer.  Note that we ignore the yellow portion of the cycle and assume that the green time is half of the light cycle.  

\begin{figure}[ht]
  \includegraphics[width=8.6cm]{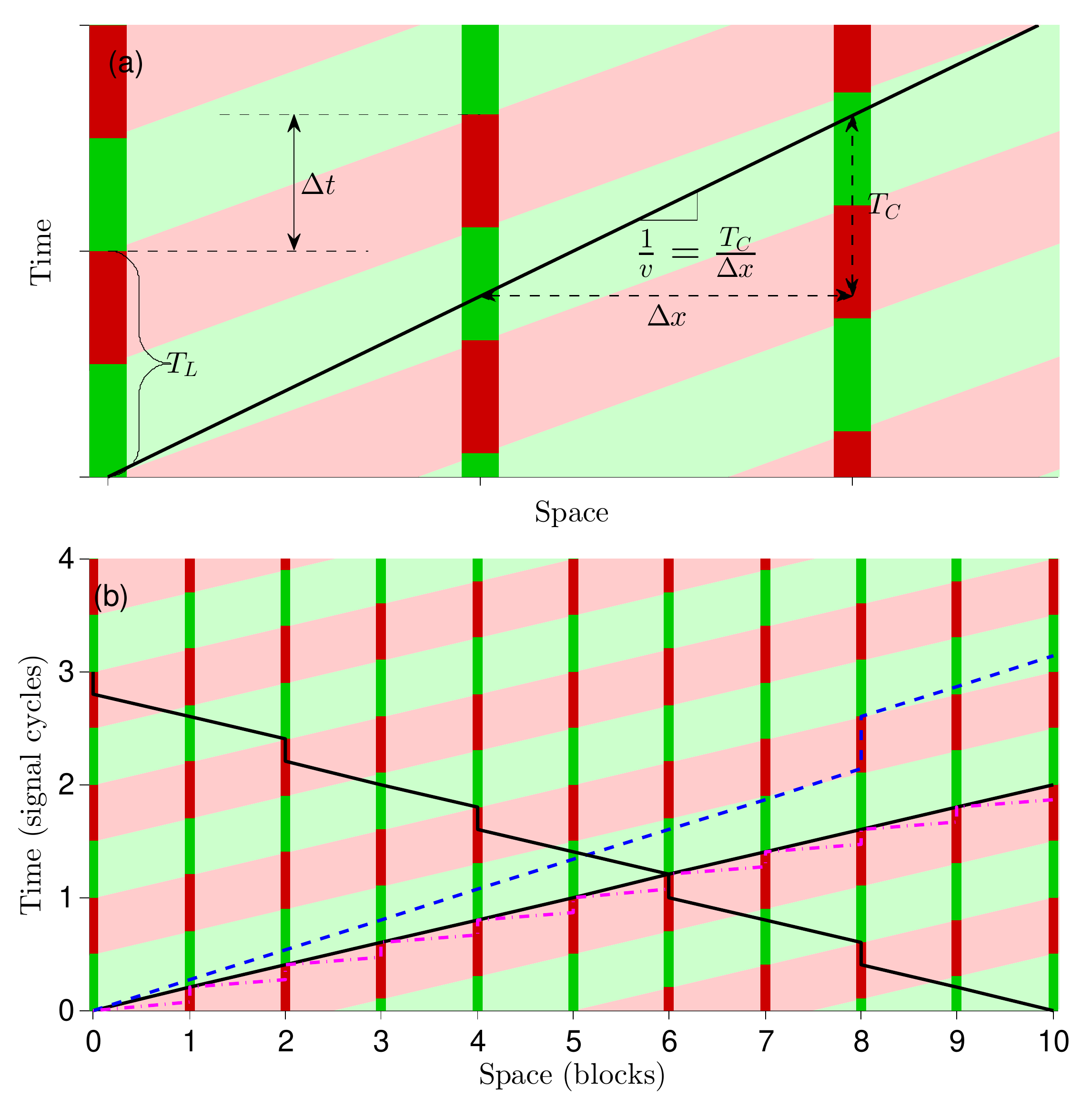}
  \caption{Space-time diagram for traffic flow.  
 Panel (a) illustrates the definitions of various variables and parameters in our model. Panel (b) displays various vehicle trajectories. The dashed line (blue) indicates a car traveling slower than the green-wave speed $v_g$, dash-dot (magenta) line indicates a car traveling faster than $v_g$, and solid lines (black) indicate cars traveling exactly at $v_g$ both eastbound (starting on the left) and westbound (starting on the right).}
  \label{fig:spacetime}
\end{figure}

Since the geometry of the system is invariant under translations of integer multiples of the block length ($ x \mapsto x+N\dx$), the optimal timing scheme should also be invariant under such translations. We therefore set
\begin{equation}
  \phi_n(t)=\frac{2\pi}{T_L} \left(t-n\dt\right)~,
\end{equation}
where $T_L$ is the period of the light cycle and $\dt$ is the time offset between consecutive lights (see Fig.~\ref{fig:spacetime}) with $0\leq\dt<T_L$. We set $\phi_0(0)=0$ without loss of generality. 

Consider a single vehicle starting at $x=0$ and traveling eastbound on this street with constant velocity $v = \dx/T_C$, where $T_C$ is the time for a car to travel one block.  When the vehicle arrives at a red light, it stops until the light turns green and then repeats the process.  From the perspective of the vehicle, at the moment this light turns green, the relative light phases are identical to the initial state.  Thus, the speed will be periodic.  

The car's effective speed $\veff$---its average speed as $t \to \infty$---is determined by the fraction of time spent waiting at red lights, suggesting that an appropriate metric for efficiency is $E = \veff/v$.  

We refer to a single cycle in which a vehicle passes through $N_L$ lights before stopping and waiting for a time $W$ as a ``trip''. During a single trip, a vehicle travels a distance of $\dx N_L$ in a total time $T_C N_L + W$.  The vehicle arrives at light $n$ at time $n T_C$, and thus $N_L$ will be the smallest positive integer satisfying 
\begin{equation}
  \label{eq:stoptime}
  (N-1/2)T_L+N_L\dt\leq N_LT_C < NT_L +N_L\dt
\end{equation}
for some integer $N$. 

Without loss of generality, we can eliminate one of the three free parameters ($T_L$, $T_C$, $\dt$) above by defining the ratios $r_C=\frac{T_C}{T_L}$ and $\rd=\frac{\dt}{T_L}$, so Eq.~\eqref{eq:stoptime} becomes 
\begin{equation}
  \label{NLeq}
  (N-1/2)\leq N_L\left(r_C-\rd\right)<N~.
\end{equation}
It is straightforward to show that $N_L=\ceil*{\frac{1}{2\{r_C-\rd\}}}$ and $N=\ceil*{N_L(r_C-\rd)}$ satisfy Eq.~\eqref{NLeq} \footnote{See Supplemental Material \S S1 for the details on solving Eq.~\eqref{NLeq}.}, where $\ceil*{x}$ and $\floor*{x}$ denote the standard ceiling and floor functions and $\{x\}=x-\floor*{x}$ denotes the fractional part of $x$ modulo 1.

The waiting time $W$ may also be computed: the car stops at time $N_LT_C$ and begins to move again at time $NT_L +N_L\dt = \ceil*{N_L(r_C-\rd)}T_L +N_L\dt$, so
\begin{equation}
	W = \ceil*{N_L(r_C-\rd)}T_L +N_L\dt -N_LT_C~.  
\end{equation}
Thus the efficiency for eastbound traffic can be expressed as   
\begin{align}
	\label{eq:eff_right}
	E_{\textrm{east}}&(r_\Delta,r_C)=\frac{r_CN_L}{\ceil*{N_L\left(r_C-\rd\right)} +\rd N_L} \nonumber \\
	&=\frac{r_C\ceil*{\frac{1}{2\{r_C-\rd\}}}}{\ceil*{\ceil*{\frac{1}{2\{r_C-\rd\}}}\left(r_C-\rd\right)} +\rd\ceil*{\frac{1}{2\{r_C-\rd\}}}}.
\end{align}
The efficiency depends on only two parameters, $r_C>0$ and $0\leq \rd<1$. It is bounded between 0 and 1, and decreases monotonically with $\rd$ except at discontinuities. It reaches a global maximum of 1 at $\rd=r_C$, which represents a green wave, and immediately after a discontinuity at $\rd=r_C+1/2$, which we refer to as a red wave. In these scenarios, lights change phase (from green to red or vice-versa) at the instant the vehicle arrives, so all waiting periods are infinitesimal.   Cross-sections of $E_{\textrm{east}}$ for fixed $r_C$ are shown in green in Figure \ref{fig:idealefficiency} \footnote{See Supplemental Material \S S2 for details about the discontinuities.}.

\begin{figure}[ht]
  \includegraphics[width=8.6cm]{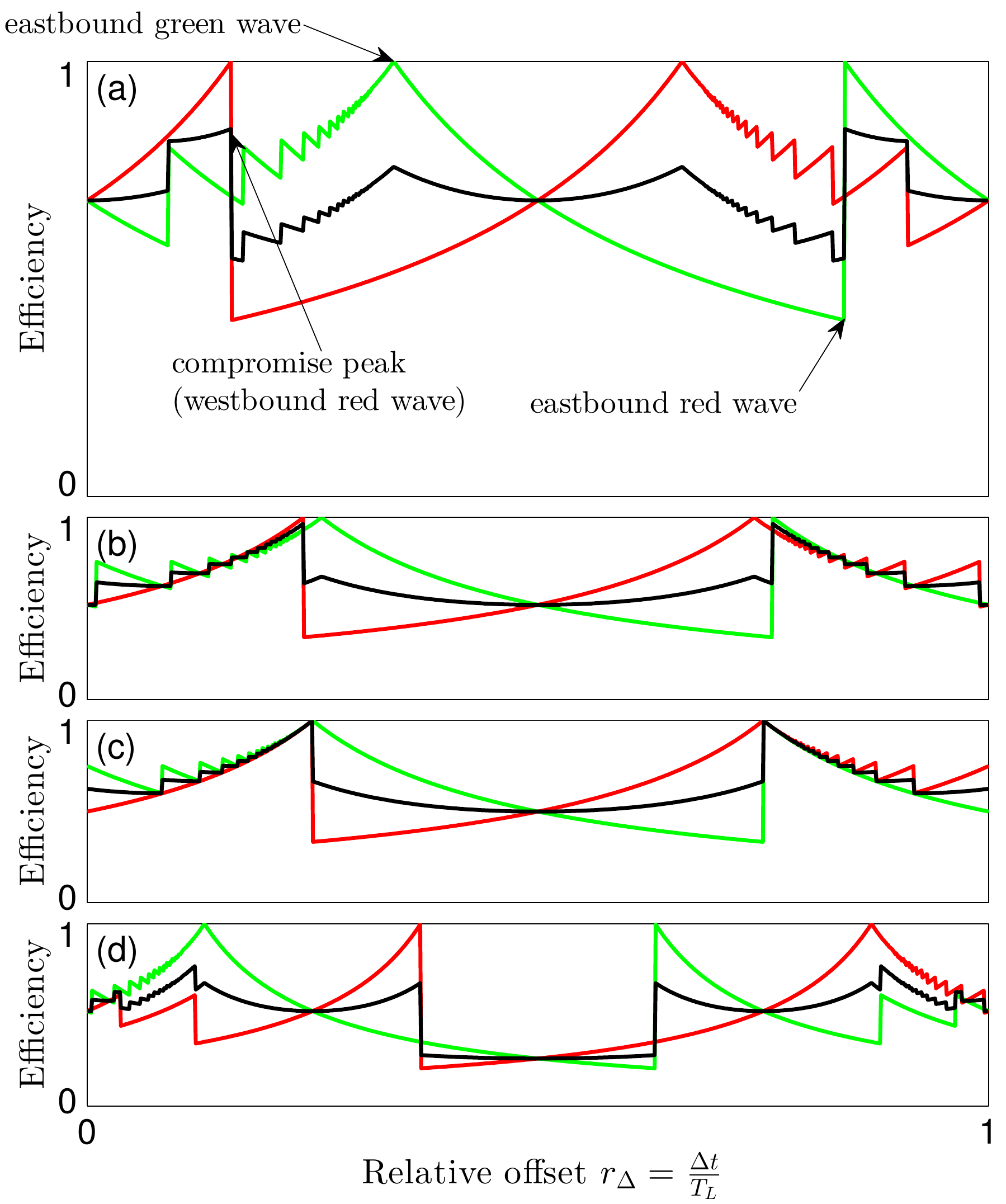}
  \caption{Theoretical efficiency versus $\rd$ for (a) $r_C=0.34$, (b) $r_C=0.26$, (c) $r_C=0.25$, and (d) $r_C=0.13$.  Green indicates $E_{\textrm{east}}$; red indicates $E_{\textrm{east}}$; and black indicates $E_{\textrm{tot}}$ assuming equal demand in both directions.}
  \label{fig:idealefficiency}
\end{figure}
 
For westbound traffic, the time delay between consecutive signals is not $\dt$ but rather $T_L - \dt$, and thus the efficiency for westbound traffic is simply $E_\textrm{west}=E_\textrm{east}(1-r_\Delta, r_C)$, the reflection of the function about $r_\Delta=0.5$.   

On a two-way street, we wish to maximize the weighted average efficiency 
\begin{equation}
  \label{eq:eff_tot}
  E_{\textrm{tot}} = w_\textrm{e} E_{\textrm{east}} + w_\textrm{w} E_{\textrm{west}}~,
\end{equation}
with weights $w_\textrm{e}$, $w_\textrm{w}$ dependent on the traffic volumes in each direction.  For simplicity we will consider the case of symmetric demand $w_\textrm{e}=w_\textrm{w}=1/2$.

Given $T_C$ as dictated by safety considerations and holding $T_L$ constant, $r_C$ is fixed, and we attempt to choose $r_\Delta$ (equivalent to choosing the offset $\dt$) to maximize efficiency.  This simultaneously maximizes the effective velocity $v_\textrm{eff}$ and minimizes the total wait time.

\begin{figure}[ht]
  \includegraphics[width=8.6cm]{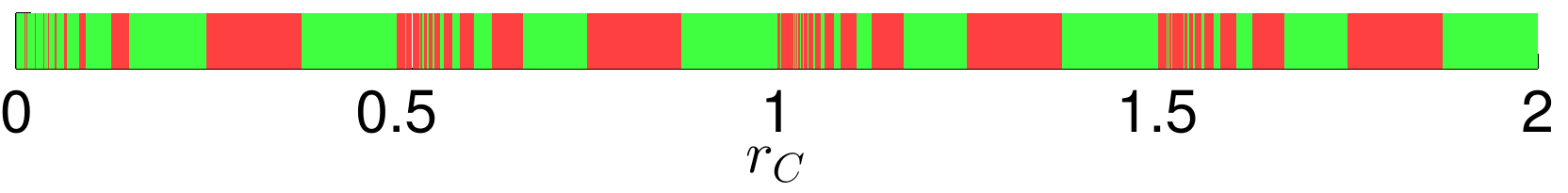}
  \caption{Optimality of the green wave.   Intervals of $r_C$ for which the green wave optimizes the bidirectional efficiency are indicated by green rectangles. Intervals of $r_C$ for which other timings are optimal are indicated by red rectangles}
  \label{fig:asymmetric}
\end{figure}

With equal demand in both directions, it might seem that the symmetry of the problem suggests a symmetrical optimum, i.e., $\rd=0$ or $\rd=1/2$ \cite{Brockfeld2001,Toledo2007}.  These are indeed local extrema, but usually not maxima. Another reasonable hypothesis is that the bidirectional optimum will coincide with the optimum in one direction, a green wave \cite{Brockfeld2001,Toledo2004,Toledo2007}.  This is a local maximum but not necessarily the global maximum.  Surprisingly, the global optimum instead often occurs when both directions are suboptimal but one direction is favored over the other.  This is possible because small perturbations in $\rd$ can cause dramatic shifts from local efficiency minima to local maxima near discontinuities in $E_{\textrm{east}}$ or $E_{\textrm{west}}$ (note that the green wave peak is not discontinuous).  When, e.g., a discontinous peak in $E_{\textrm{west}}$ occurs near the green wave peak for $E_{\textrm{east}}$,  the loss of efficiency by perturbing off of the eastbound green-wave is offset by gains in the westbound efficiency.  As a result, green wave timings fail to be optimal for various ranges of $r_C$ (see Fig.~\ref{fig:asymmetric}).

\PRLsec{Some limitations}While the efficiency metric in Eq.~\eqref{eq:eff_tot} provides some insight into the ideal signal timing for a two way street, it has a number of limitations.  First of all, it applies only to a single vehicle.  In practice, vehicles often travel in groups known as platoons \cite{Newell1980,Gartner1991}.  The jagged efficiency peaks described by Eq.~\eqref{eq:eff_right} and displayed in Fig.~\ref{fig:idealefficiency} may not be achievable by an entire platoon of vehicles. The theory also assumes identical non-interacting cars with constant speeds and perfectly uniform light spacing.  In practice, city blocks vary in length even in well-planned urban grids and driver behavior varies. Additionally, the interactions between vehicles can play a significant role in exacerbating congestion \cite{Huang2003}.

\begin{figure}[ht]
  \includegraphics[width=8.6cm]{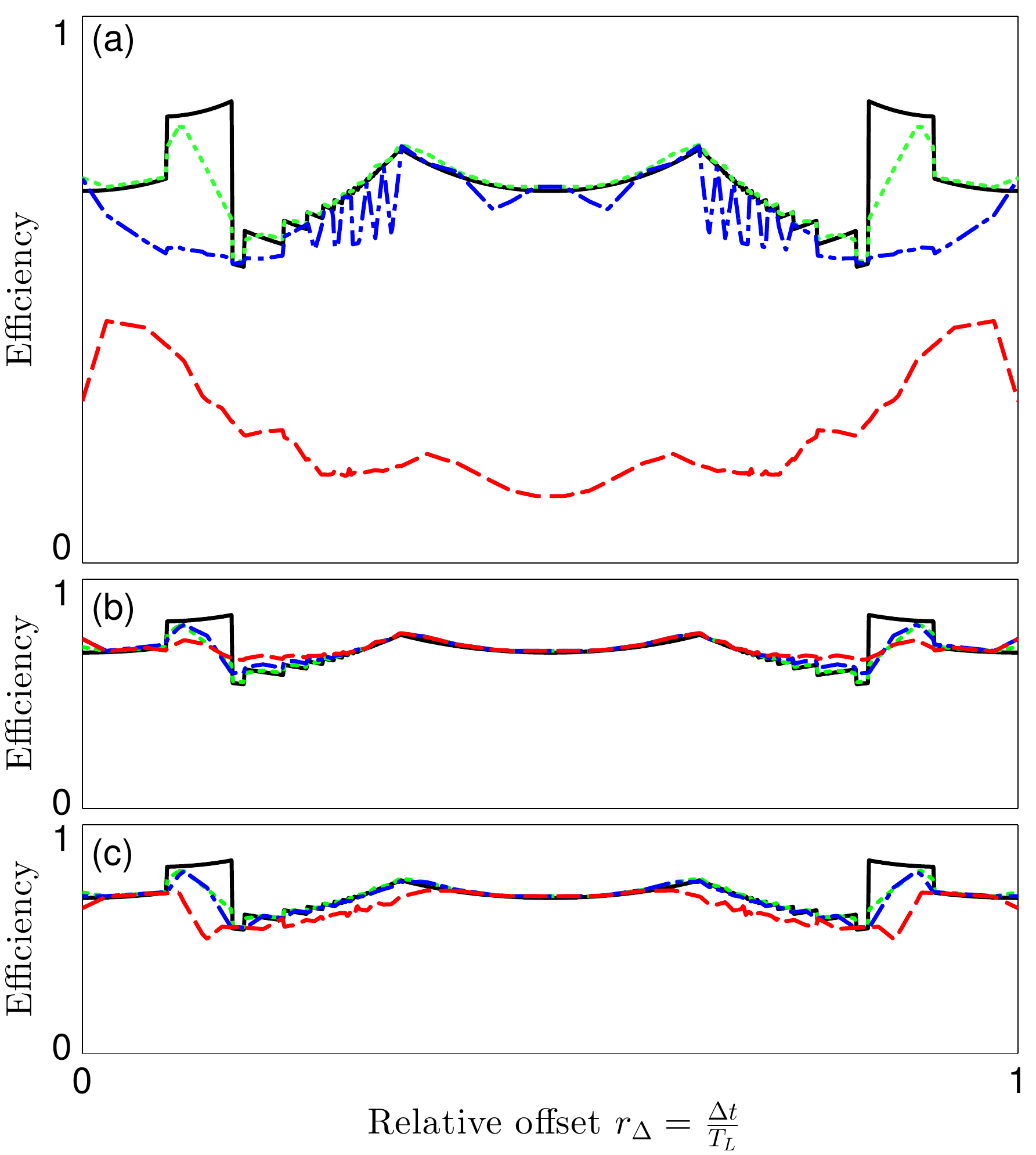}
  \caption{Efficiency from simulation versus $r_\Delta$ for $r_C=0.34$.   Panel (a) shows the efficiency for increasing vehicle density: the solid curve (black) indicates the theoretical efficiency $(E_{\textrm{tot}})$; the dotted (green), dash-dotted  (blue), and dashed (red) curves represent simulation results for vehicle densities of 10\%, 50\%, and 90\% density respectively. The bottom panels display the effects of variation in the traffic signal spacing (b) and vehicle speed (c) on the efficiency with 10\% traffic density. Solid (black) indicates the theoretical efficiency; the dotted (green), dash-dotted  (blue), and dashed (red) curves represent simulation results for  0\%, 1\% and 5\% standard deviation respectively.}
  \label{fig:sim_efficiency}
\end{figure}

To test the predictions of our model and verify that they are relevant when these assumptions are relaxed, we simulated the flow of vehicles on a street with fifty periodically placed traffic lights.  We imposed periodic boundary conditions to avoid arbitrarily specifying entrance and exit rates.  Instead, cars were randomly placed along the street according to a specified density $\rho$ representing the fraction of the system occupied by vehicles of a finite length, and the total number of vehicles in the system was conserved. The trips of these vehicles were simulated during 30 light cycles.  In the simulation, vehicles were prevented from passing each other. This caused queues to form at red lights as one might expect. Simulations were repeated with unevenly spaced lights and variable vehicle speed; results can be found in Fig.~\ref{fig:sim_efficiency}.

\PRLsec{Simulation results}Simulations reveal that at low vehicle densities, the computed efficiency is well-approximated by the theory and non-green-wave optima persist (see panel (a) of Fig.~\ref{fig:sim_efficiency}).  At moderate densities, the efficiency near discontinuous peaks degrades noticeably while the green wave remains highly efficient. Thus a perfect green wave in either direction is optimal for moderate densities. At very high densities, gridlock, the scenario where vehicles at green lights are unable to advance due to the queue ahead of them, becomes a significant issue and the efficiency of all timings degrades.  The only way to avoid gridlock is to set $\rd=0$ and have all lights change in unison.  

The middle and bottom panels of Fig.~\ref{fig:sim_efficiency} show the efficiency when vehicle speed varies (panel (b)) and when the light spacing varies (panel (c)). Variation in the light spacing with proportionate variation in the offsets can actually improve efficiency for some ranges of $\rd$.  This is reasonable given that lights that are close together behave like a single light and lights that are far apart have smaller wait times relative to the travel times.  Variation in the vehicle speed has a smoothing effect on the discontinuities in the efficiency curve.  Both of these factors degrade the efficiency in a smooth way, allowing the discontinuous optima to persist when the variation is small.  Thus theoretical predictions are ``structurally stable'' in the sense that they may indeed have value even in real-world systems with non-ideal behavior. 

\PRLsec{Density effects}To explore the effects of vehicle density in greater detail, we computed the efficiency for fixed $(\rd,r_C)$ and increasing $\rho$.  Below a critical density, which we refer to as the ``jamming threshold'' \cite{Sasaki2003}, the predictions of Eq.~\eqref{eq:eff_tot} give a good approximation for the efficiency.  Above this threshold, the efficiency degrades, and the theory no longer approximates the observed behavior. For a range of physically relevant parameters the critical density is above 50\% of the capacity of the road.  Near some discontinuous peaks, however, the critical density is small and few vehicles are able to perform at the level indicated by the theory.  This is due to the narrow corresponding to these timings \footnote{See Supplemental Material \S S3 for the derivation of the bandwidth.}. Nonetheless, discontinuous peaks in the bidirectional efficiency can remain optimal for a range of densities beyond the threshold.

\begin{figure}[ht]
  \includegraphics[width=8.6cm]{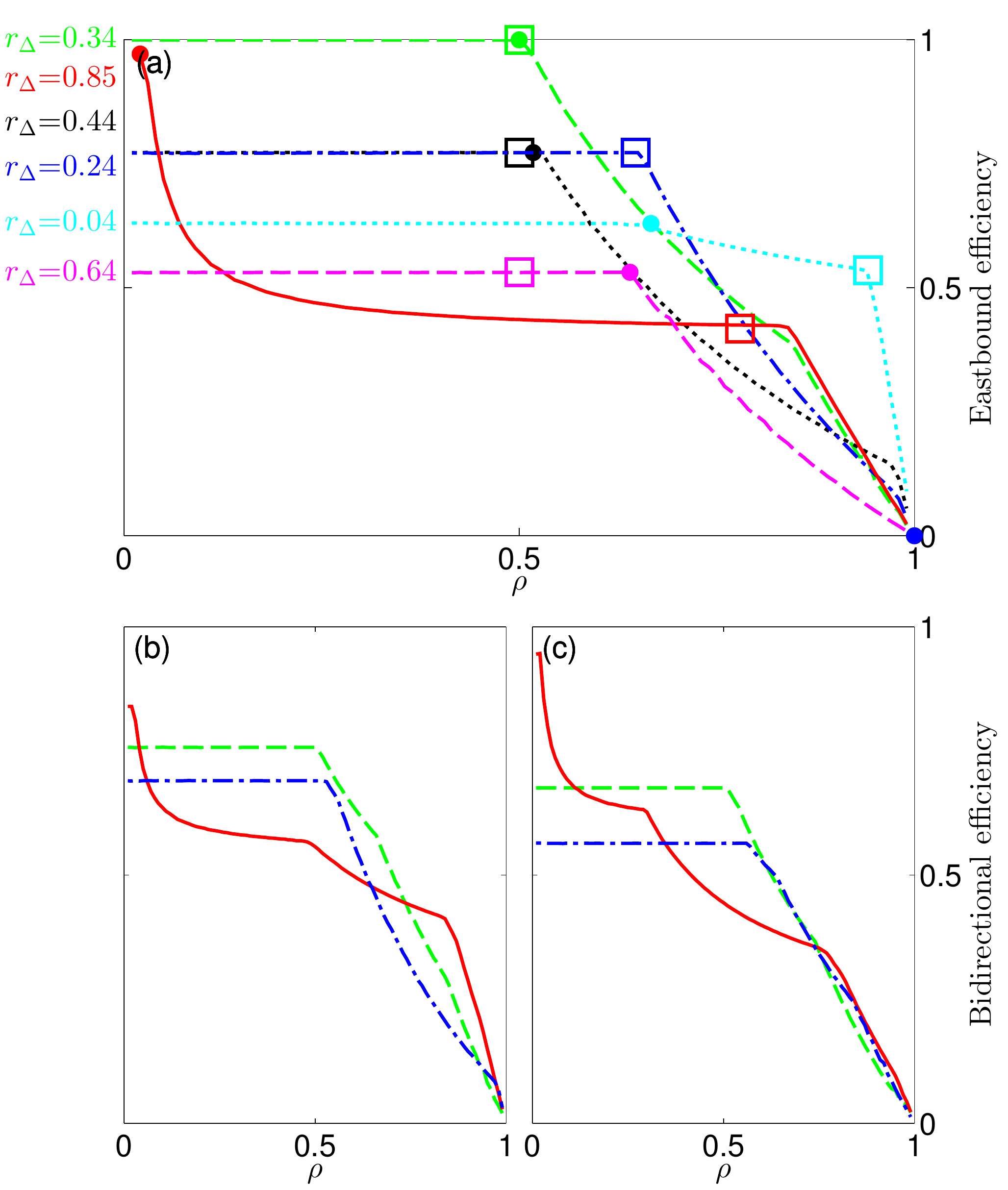}
  \caption{Predictions and simulations of various jamming transitions. Panel (a) displays the eastbound efficiency for various values of $\rd$ with $r_C=0.34$.  Of particular interest are the green curve (dashed), which represents an eastbound green wave, and the red curve (solid), which is near an eastbound red wave.  The markers correspond to the critical densities for platoon segmentation (circle) and platoon coalescence (square). Panels (b) and (c) show the bidirectional (weighted average of east and westbound) efficiency for $r_C=0.34$ and $r_C=0.26$ respectively with curves representing green wave peaks (green dashed curve), discontinuous peaks (red solid curve), and suboptimal timings (blue dash-dot curve).}
  \label{fig:jamming}
\end{figure}

The degradation of the quality of the theoretical predictions is due to the assumption that vehicles are non-interacting.  Above the jamming threshold, the interactions between vehicles cause delays that the theory ignores.  In simulations, vehicles initially clump together forming platoons.  These platoons can interact either by coalescing to form even larger platoons or by being segmented at red lights.  We can estimate the critical density corresponding to the jamming threshold analytically by deriving the conditions under which this coalescence and segmentation occurs at steady state \footnote{See Supplemental Material \S S4 for the equation and derivation of the jamming threshold and \S S5 for videos of the platoon dynamics.}.  These predictions are displayed along with the numerical results in Fig.~\ref{fig:jamming}.    
 
\PRLsec{Conclusions}In the mid-20th century physicists and engineers began taking an analytical approach to traffic management. Theoretical work has since proceeded along several lines, but we believe that there is still insight to be gained from simple solvable models.  

Here we have presented a new analysis of traffic flow on an urban arterial road with periodic traffic signals.  Our approach allows analytical prediction of optimal signal timing that agrees well with numerical simulation and approximates the behavior of the system even when idealizing assumptions are relaxed.  It yields an efficiency metric that can be expressed and computed analytically, yet reproduces features observed in more complex models---features such as platoon formation \cite{Nagatani2006}, discontinuous efficiency curves \cite{Toledo2007,Huang2003}, irregular flow patterns \cite{Toledo2004}, and phase transitions due to jamming \cite{Sasaki2003,Huang2003,Lighthill1955}.  

These findings should provide useful insight for the design of coordinated traffic signal timing programs on long arterials, particularly during periods of low to moderate traffic demand.  The model could be used to generate initial guesses for numerical optimization schemes with more complex efficiency metrics, or, alternatively, our efficiency metric \eqref{eq:eff_tot} could be modified to apply to arbitrary networks of one- and two-way streets, perhaps allowing exact rather than approximate optimization and yielding more intuitive understanding of results.

\begin{acknowledgments}
  The authors thank H. Mahmassani and A. Motter for useful discussions.  Research was supported in part by a NICO seed grant.
\end{acknowledgments}

\end{document}


\title{Supplemental Material: Symmetry breaking in optimal timing of traffic signals on an idealized two-way street} 

\author{Mark J. Panaggio}
\email[email: ]{markpanaggio2014@u.northwestern.edu}
\affiliation{Department of Engineering Sciences and Applied Mathematics, Northwestern University, Evanston, IL 60208, USA}

\author{Bertand J. Ottino-L\"offler}
\affiliation{Department of Mathematics, Caltech, Pasadena, CA 91126, USA}

\author{Peiguang Hu}
\affiliation{Engineering Science Program, National University of Singapore, Singapore}

\author{Daniel M. Abrams}
\affiliation{Department of Engineering Sciences and Applied Mathematics, Northwestern University, Evanston, IL 60208, USA}
\affiliation{Northwestern Institute on Complex Systems, Northwestern University, Evanston, IL 60208, USA}
\date{ \today}

\maketitle
\section{Derivation of eastbound efficiency}
Recall that the length of a trip is determined by $N_L$ (a trip represents the periodic trajectory in which a vehicle passes through $N_L$ lights before stopping and waiting for a time $W$) which satisfies the following.
\begin{equation}
  \label{NLeq}
  (N-1/2)\leq N_L\left(r_C-r_{\Delta}\right)<N~.
\end{equation}
For simplicity, we define $M=r_C-\rd$. We can eliminate $N$ from the above expression by noting that it is the smallest integer that satisfies 
\begin{align*}
  N_LM<N\leq N_LM+\frac{1}{2}
\end{align*}
for a positive integer $N_L$.  Clearly this is satisfied if $N=\ceil{N_LM}$ where $N_L$ satisfies the equation
\begin{equation*}
  \label{NLeq_supp}
  \floor*{N_LM+\frac{1}{2}}=\ceil{N_LM}
\end{equation*}
This equation is periodic in M.  This can be verified by noting that we can write $M=\floor*{M}+\{M\}$.  Expanding the arguments of both sides of the equation and simplifying integer terms, we note that the resulting equation depends only on the fractional part $\{M\}$    
\begin{equation}
\label{NLeq_f}
\floor*{N_L\{M\}+\frac{1}{2}}=\ceil{N_L\{M\}}
\end{equation}
It is straightforward to show that the solution to this equation is $N_L=\ceil{\frac{1}{2\{M\}}}$.  Suppose $N_L<\frac{1}{2\{M\}}$.  Then 
\begin{align*}
	\frac{1}{2}\leq N_L\{M\}+\frac{1}{2}<\frac{1}{2\{M\}}\{M\}+\frac{1}{2}<1&\implies \floor*{N_L\{M\}+\frac{1}{2}}=0 \textnormal{ and } \\
	0\leq N_L\{M\}<\frac{1}{2\{M\}}\{M\}<1/2&\implies \ceil{N_L\{M\}}=1.
\end{align*}
So $N_L<\frac{1}{2\{M\}}$ cannot be possible.  Now suppose $N_L=\ceil{\frac{1}{2\{M\}}}$ (the smallest integer greater than or equal to $\frac{1}{2\{M\}}$).  If $\{M\}>1/2$, then $N_L=1$ satisfies the equation, so the assumption is satisfied.  If $\{M\}\leq 1/2$, then 
\begin{align*}
	1=\frac{1}{2\{M\}}\{M\}+\frac{1}{2}\leq N_L\{M\}+\frac{1}{2}<\left(\frac{1}{2\{M\}}+1\right)\{M\}+\frac{1}{2}<1+\{M\}&\implies \floor*{N_L\{M\}+\frac{1}{2}}=1 \textnormal{ and }\\
	1/2= \left(\frac{1}{2\{M\}}\right)\{M\}\leq N_L\{M\}<\left(\frac{1}{2\{M\}}+1\right)\{M\}<\frac{1}{2}+\{M\}<1
	&\implies \ceil{N_L\{M\}}=1.
\end{align*}
So, $N_L=\ceil{\frac{1}{2\{M\}}}$ must be the smallest integer solution to Eq.~\eqref{NLeq_f}. 
\section{Exploration of discontinuities}

In our model, an isolated vehicle always travels an integer number of blocks before stopping.  This results in a discontinuous eastbound efficiency.  For $r_C=\rd$, $N_L$ is infinite. For $\rd$ slightly greater than $r_C$, $N_L$ drops to 1 and then increases for increasings $\rd$ (mod 1).  $N_L$ is always an integer, so it increases in integer increments, and each jump produces a jump in efficiency.  Immediately after the jump, the efficiency reaches a local maximum, before decreasing again.  It can be shown that these peaks in efficiency are located at the following locations:
\begin{equation}
\label{peak_locs}
	\rd-r_C=\left\{0, 1/2,3/4, ...\frac{2N_L-1}{2N_L}\right\}. 
\end{equation} 
 
The first peak $\rd=r_C$ represents a green wave, the ideal case in which all vehicles travelling in a single direction are able to proceed indefinitely without stopping.  The second peak corresponds to timings in which a vehicle leaving a green light at the instant it turns green arrives at the next light immediately before it turns red.  Thus the car travels two blocks and then stops for an instant at a red light before it turns green.  We call this a ``red wave'' because the absolute minimum efficiency(where cars stop at every red light and wait for the red cycle) lies immediately to the left of this peak.  The third peak occurs when the car makes it through the first two lights (arriving at the second at the instant before it turns red), but then stops at the third and waits the remainder of the red cycle.  Subsequent peaks are analogous.

\begin{figure}[ht]
  \includegraphics[width=17.2cm]{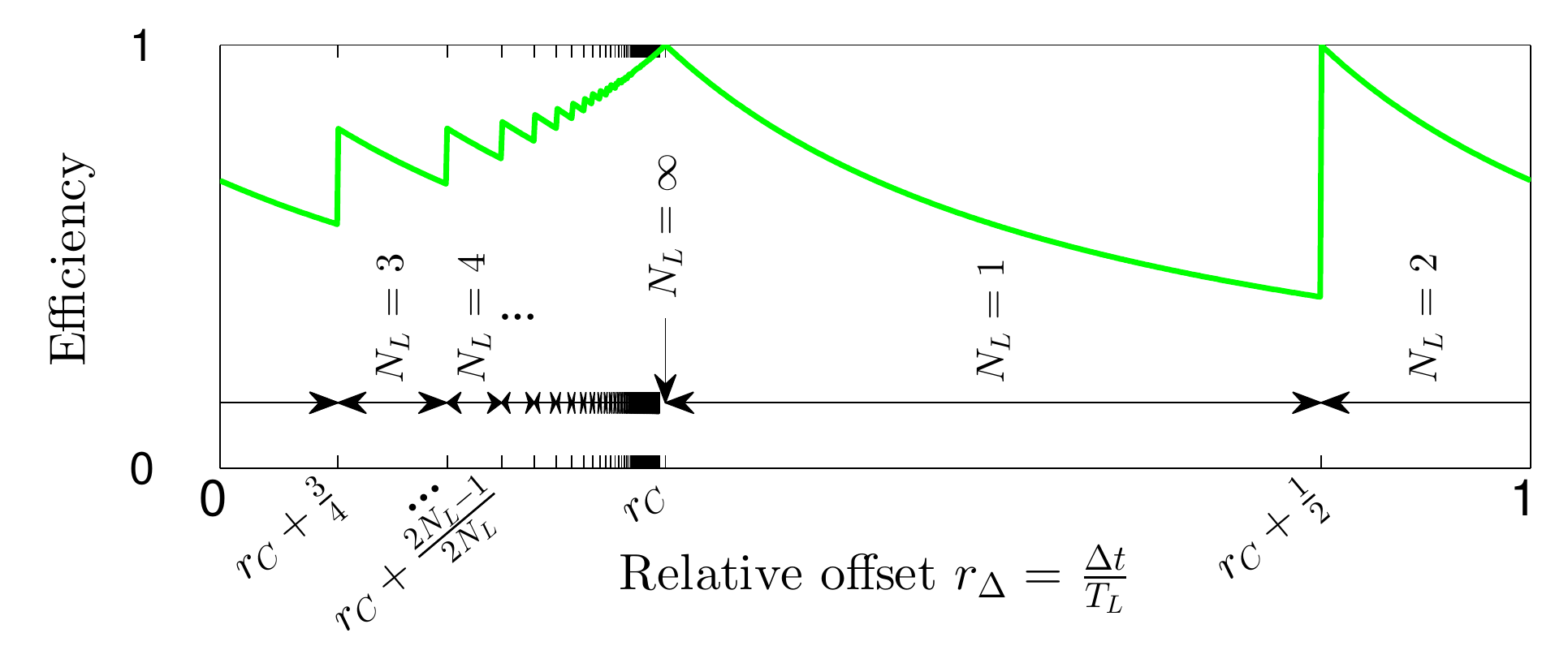}
  \caption{Peaks of theoretical efficiency. The eastbound efficiency is plotted for $r_C=0.34$ with ticks marking the local efficiency maxima. }
  \label{fig:peaks}
\end{figure}

Between these peaks, the efficiency decays like
\begin{displaymath}
   E = \left\{
     \begin{array}{lr}
		\frac{N_Lr_C}{N_L\rd+1} & \textnormal{ for $\rd<r_C$} \\\\
		\frac{N_Lr_C}{N_L\left(\rd-1\right)+1} & \textnormal{ for $\rd>r_C$}
     \end{array}
   \right.
\end{displaymath} 
reaching minima of 
\begin{align*}
	E&=\frac{N_Lr_C}{N_Lr_C+1/2} 
\end{align*} 
and maxima of 
\begin{align*}
	E&=\frac{N_Lr_C}{N_Lr_C+1 -\frac{N_L}{2(N_L-1)}}. 
\end{align*}
This is illustrated in Fig.~\ref{fig:peaks}.
  
\section{Computation of bandwidth}
In this system, each intersection has a maximum flow rate that is determined by the vehicle speed and the green time. This flow rate limits the length of a platoon that can clear the light in a single cycle.  We refer to this as $L_0$.  
\[L_0=v\frac{T_L}{2}=\frac{\Delta x T_L}{2T_C}=\frac{\Delta x}{2r_C}\]
Although a platoon of length $L_0$ is able to clear a single light during one cycle, the platoon may be unable to clear subsequent lights.  The theory in the manuscript is only directly applicable to a single vehicle. It is possible to show that there is a platoon length $L$ greater than the length of a single vehicle for which this theory is also valid.  We define bandwidth $B=\frac{L}{L_0}$ as the fraction of the maximum allowable platoon length that can achieve the efficiency from theory.  This can also be interpreted as the fraction of departures during the green time that will allow vehicles to achieve the theoretical efficiency.  This is simply the standard definition of bandwidth in seconds normalized by the green time $T_L/2$.   Thus a bandwidth of 1 means that a platoon of length $L_0$ is able to achieve the efficiency of a single vehicle.  A bandwidth of 0 means that only the single car can achieve this efficiency, and that no other vehicles can perform as well.

There are two factors that restrict the bandwidth: the phases of downstream lights upon the arrival of the platoon and the initial phases of upstream lights.  

\PRLsec{Downstream bandwidth}To compute the downstream bandwidth, we note that the first vehicle in a platoon arrives at light $n$ at time $nT_C$.  For lights $1,2,...,N_L-1$, the light is green on arrival and the remaining green time restricts the bandwidth.  The phase of light $n$ upon arrival is $n\omega(T_C-\dt)$ which satisfies \[k_n2\pi<n\omega(T_C-\dt)<(k_n+1/2)2\pi\] for integer $k_n=\floor*{n(r_C-\rd)}$ where the leftmost expression represents the phase when the light turns green and the rightmost expression represents the phase when the light turns red. The remaining green time is \[\left(k_n+\frac{1}{2}\right)T_L - n(T_C-\dt).\]  We divide this by the total green time $T_L/2$ for proper normalization.  Taking the minimum over the first $N_L-1$ lights yields the downstream bandwidth
\begin{equation*}
	B_{\textnormal{down}}=2\min\{\floor*{n(r_C-\rd)}+1/2-n(r_C-\rd)\}_{n=1}^{N_L-1}.
\end{equation*}
Note that for $r_C<\rd<r_C+1/2$, the downstream bandwidth is 1.  This is due to the fact that vehicles encounter a red after traveling a single block.   In this region, the upstream lights limit the bandwidth.

\PRLsec{Upstream bandwidth}The upstream bandwidth is necessary because $L_0$ may exceed the length of a single block allowing an upstream light to segment the platoon. Given a platoon of length $L_0$ waiting at light $0$, we now determine which of those vehicles will be able to reach light 0 before it turns red.  Upstream lights -1 to $-n$ are green if $n\dt<T_L/2$ or $n<\frac{1}{2\rd}$.    We choose $n$ as the total number of consecutive upstream green lights 
\[n=\floor*{\frac{1}{2\rd}}.\]
This limits the length of platoon that can clear light 0 to at most $n+1$ blocks.  The remaining green time for light $-n$ further restricts the length of the last fragment.  It remains green for time $\frac{T_L}{2}-n\dt$, so only vehicles within $v\left(\frac{T_L}{2}-n\dt\right)$ of light $-n$ will be able to advance before it turns red.  As a result, the platoon fragment beyond light $-n$ is limited to the lesser of $v\left(\frac{T_L}{2}-n\dt\right)$ and $\Delta x$.  Hence the upstream bandwidth is
\begin{align*}
	B_{\textnormal{up}}&=\frac{1}{L_0}\min\left(L_0, n\Delta x+\min\left[\Delta x,v\left(\frac{T_L}{2}-n\dt\right)\right]\right)\\
	&=\min\left(1, \floor*{\frac{1}{2\rd}}2r_C+\min\left[2r_C,\left(1-\floor*{\frac{1}{2\rd}}2\rd\right)\right]\right).
\end{align*}

Note that for slow vehicles ($r_C\geq1/2$) the upstream bandwidth is 1; this occurs because $L_0$ is less than one block making the phases of upstream lights irrelevant.  For very fast cars ($r_C\ll 1/2$), $L_0$ is large, and the upstream lights tend to play the dominant role in determining the bandwidth. 

\begin{figure}[ht]
  \includegraphics[width=8.6cm]{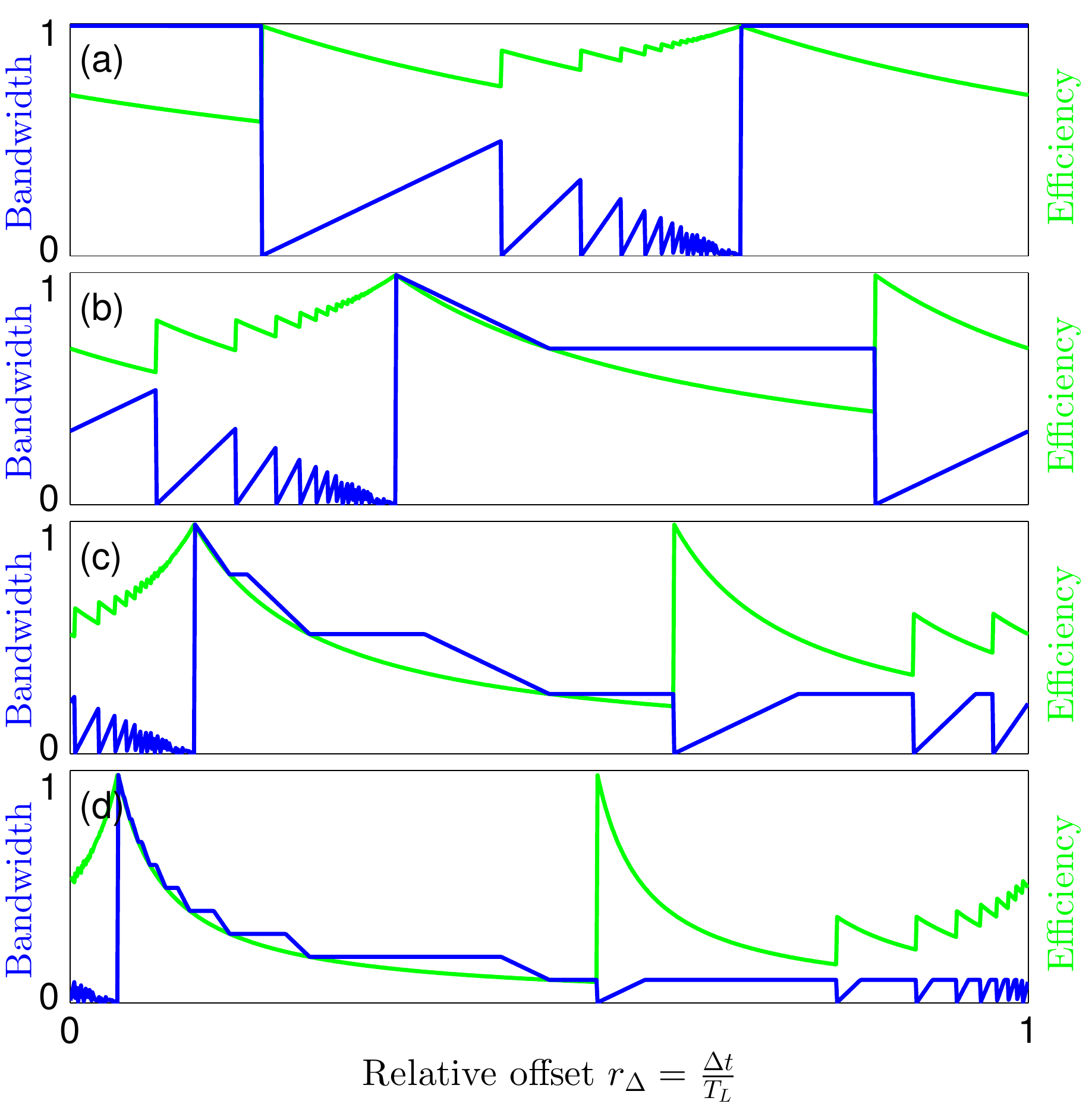}
  \caption{Bandwidth of theoretical efficiency. The bandwidth (blue) as a function of $\rd$ is displayed along with the eastbound efficiency (green) for (a)  $r_C=0.7$, (a)  $r_C=0.34$, (a)  $r_C=0.13$, and (a)  $r_C=0.05$. }
  \label{fig:bw}
\end{figure}

\PRLsec{Composite bandwidth}
The true bandwidth is the minimum of the upstream and downstream bandwidths
\begin{equation}
	B=\min\left(B_{\textnormal{down}},B_{\textnormal{up}}\right),
\end{equation} 
and it is bounded $0\leq B \leq1$. Figure \ref{fig:bw} overlays the bandwidth on the plot of the eastbound efficiency.  It is clear that all peaks with the exception of the green wave peak have zero bandwidth. 

One shortcoming of this approach is that it does not provide information about how the other vehicles perform relative to the first car in the platoon.  Thus small bandwidth cannot necessarily be equated with poor efficiency or even large deviation from the theoretical efficiency. It simply indicates that deviations from the theory are possible. Large bandwidth, on the other hand, indicates that the theory is likely valid even for moderate traffic densities. Simulations reveal that despite the low bandwidth, timings near secondary peaks still perform well at low densities.   

\section{Jamming threshold}
In our simulations, the cars were randomly placed in the system at according to an initial density $\rho$. We observed that, below a critical density $\rho_{\textnormal{crit}}$, the efficiency is independent of density and is accurately predicted by the theory. Above that density, the efficiency decreases as vehicle density increases.  In the theory, we assume vehicles are non-interacting.  In simulations with large numbers of vehicles, this assumption no longer holds.  Thus the degradation of efficiency can be attributed to the increasing frequency of interactions between vehicles.  We now compute the ``Jamming threshold'', the critical vehicle density for which vehicle interactions can no longer be ignored. 

Vehicles interact when they stop behind other vehicles waiting at red lights.  At low densities, this causes vehicles to form platoons which then behave as a single unit.  These interactions cease once a steady state platoon distribution is reached, and consequently they do not cause significant reductions in efficiency. However, when these platoons exceed a critical length, either repeated segmentation by red lights or coalescence with other platoons becomes unavoidable.   When these interactions cause vehicles to stop earlier than they would have in isolation, the efficiency decreases. 

The two types of platoon interactions that cause a degradation of efficiency, platoon coalescence and platoon segmentation, become significant at different densities.  So we compute two different thresholds that can explain the behavior visible in Fig.~5. 

\begin{figure}[ht]
  \includegraphics[width=17.2cm]{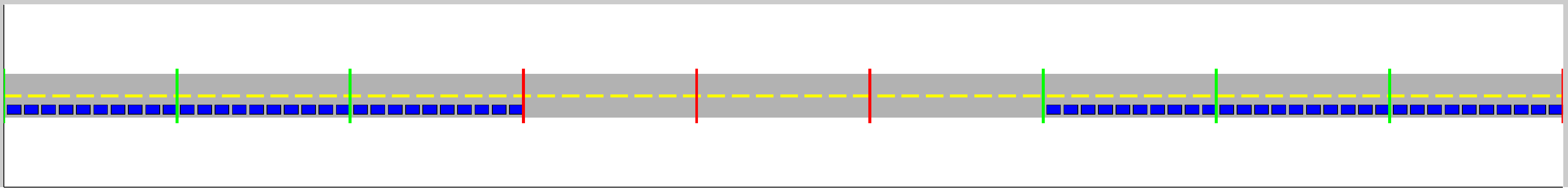}
  \caption{Platoon distribution near the jamming threshold for $r_C<\rd$. This displays a snapshot from a simulation for $\rd=1/6$ where eastbound cars travel faster than the green wave.  The green and red vertical bars represent the traffic lights, and the the blue rectangles represent individual vehicles.  This snapshot is taken at the instant before the light on the far right turns green and the light at the trailing end of the platoon turns red.  Note that for this particular value of $\rd$ the green string consists of $n=\floor*{\frac{1}{2\rd}}=3$ lights. This string is filled with a platoon of vehicles that advances at the green wave speed.  Adding additional vehicles would cause vehicles at the trailing end of the green string to stop prematurely due to the coalescence at the front of the platoon.}
  \label{fig:fastcars}
\end{figure}

\PRLsec{Platoon coalescence}The frequency of coalescence depends on the relative speeds of the green wave and the individual vehicles.   When $0\leq \dt <\frac{T_L}{2}$ and $\dt > T_C$, cars travel faster than the green wave. As a result, vehicles accumulate at the front of each string of green lights. After the initial transients decay, these vehicles form platoons that behave as a unit provided the platoon does not fill up the series of consecutive green lights. See Fig.~\ref{fig:fastcars} for an illustration of this scenario. Since half of the system is occupied by green lights and there are no vehicles between red lights at equilibrium, the critical density must be $\rho_{\textnormal{crit}}=\min\left[1/(2r_C),1/2\right]$.

\begin{figure}[ht]
  \includegraphics[width=17.2cm]{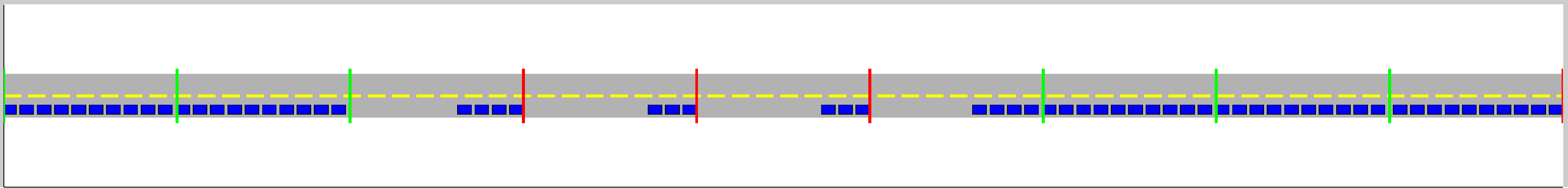}
  \caption{Platoon distribution near the jamming threshold for $r_C>\rd$. This displays a snapshot from a simulation for $\rd=1/6$ where eastbound cars travel slower than the green wave.  This snapshot is taken at the instant before the light on the far right turns green and the light at the trailing end of the long platoon in the green string  turns red.  Note that for this particular value of $\rd$ the green string is filled with vehicles while shorter platoons remain queued at red lights. Adding additional vehicles would cause the short platoons behind red lights to increase in length thereby causing vehicles at the end of the long platoon to stop prematurely when coalescence takes place.}
  \label{fig:slowcars}
\end{figure}

When $0\leq \dt <\frac{T_L}{2}$ and $\dt \leq T_C$, cars travel slower than the green wave. Thus, the front end of the red wave catches vehicles in the green section at a constant rate. This leaves queues of constant length in the red section which are then left behind at the trailing end of the red wave. However, if there are too many vehicles in the green section, then the cars at the front of the green section will be slowed down by the cars left behind by the red wave.  This occurs when the green section is full of cars (density = 1), and the platoons in the red section surpass a critical length determined by the ``catch up rate'' (density = $1-\frac{\rd}{r_C}$). See Fig.~\ref{fig:slowcars} for an illustration of this scenario. Thus the critical density is a weighted average of the densities in the green segments and red segments: $\rho_{\textnormal{crit}} =  1/2+1/2\left(1-\rd/r_C\right)$.

For left moving waves ($\rd>1/2$) the behavior is more complex and less intuitive.  This case is less relevant for urban networks where $r_C$ will typically be small (optimal $\rd$ are generally close to $r_C$).  Nonetheless, replacing $\rd$ with  $1-\rd$ in the solution for right-moving waves yields good agreement with simulations. 

The predicted thresholds for platoon coalescence are marked by an open square and displayed in panel (a) of Fig.~5 in the main text. 

\PRLsec{Platoon segmentation} Platoon segmentation is related to the bandwidth.  A platoon longer than the bandwidth will necessarily be segmented as it travels through the system.  At the threshold, the system is filled with platoons of length $L_P=L_0B$.  These platoons are separated by gaps of length $G$.  Simulations suggest that the variation in $G$ is small near the threshold, so we assume $G$ is a constant.   Under this assumption, the critical density of vehicles in the system is 
\begin{equation}
\label{rhoc_eq1}
 \rho_{\textnormal{crit}} = \frac{L_P}{L_P+G}
\end{equation}
In order to determine this density, one additional equation is needed.  A platoon of length $L_0$ will be segmented into $N_L$ pieces by downstream lights. Thus the region of length $L_0$ will contain $N_L$ platoons and $N_L-1$ gaps. The missing equation is 
\begin{equation}
\label{rhoc_eq2a}
 N_LL_P+(N_L-1)G=L_0 \textnormal{ for } N_L>1. 
\end{equation}
If $N_L=1$, this equation is redundant and instead 
\begin{equation}
\label{rhoc_eq2b}
 L_P+G=\Delta x/\rd \textnormal{ for } N_L=1. 
\end{equation}
This is due to the fact that there is one platoon per ``wavelength'' in the green wave. Equations \eqref{rhoc_eq1}, \eqref{rhoc_eq2a} and \eqref{rhoc_eq2b} can be solved for $\rho_{\textnormal{crit}}$ yielding the predictions marked by a solid circle and displayed in panel (a) of Fig.~5 in the main text. 

\section{Supplemental videos}
Three video clips from simulations are included to demonstrate the platoon dynamics observed.  These display eastbound and westbound traffic on an 8 block segment of the total 50 block long system.  Traffic lights are indicated by red and green vertical lines, and individual vehicles are marked by blue rectangles.  The videos included correspond to $r_C=0.34$, $\rho=0.25$, and three different values of $\rd$ corresponding to a non-green-wave optimum ($\rd=0.15$), a green wave ($\rd=0.34$), and a sub-optimal timing ($\rd=0.44$).  The total inflow and outflow for both eastbound and westbound traffic are also displayed.  Due to the short timespan over which these totals are computed, they are highly dependent on the initial condition and as a result should not be viewed as a reflection of effectiveness of the timing scheme.  
\bibliography{TrafficCitations}		

%